\documentstyle[11pt]{article}
\newlength{\pubnumber} 

\catcode`\@=11
\@addtoreset{equation}{section}

\def\section{\@startsection{section}{1}{\z@}{3.5ex plus 1ex minus .2ex}
{2.3ex plus .2ex}{\large\bf}}
\def\subsection{\@startsection{subsection}{2}{\z@}{2.3ex plus .2ex}
{2.3ex plus .2ex}{\bf}}

\def\elxpsl{\hbox{/\kern-.4500em$\ell$}}

\def\lt{{\tilde{\lambda}}}
\def\la{\lambda}
\def\A{{\cal A}}
\def\L{{\cal L}}
\def\M{{\cal M}}

\begin{document}

\begin{titlepage}

\samepage{
\setcounter{page}{1}
\vskip 5.5truecm
\begin{center}
\renewcommand{\thefootnote}{\fnsymbol{footnote}}
\setcounter{footnote}{0}
\vskip 9truecm

{\large\bf More on Four-Dimensional Extremal Black Holes\\}
  
 \vskip 2truecm
     {\bf N.~ Hambli}
\footnote{E-mail address: hambli@hep.physics.mcgill.ca or 
hambli@lpsvsh.lps.umontreal.ca.}
\footnote{Address after September 1996: Institut 
des Hautes \'Etudes Scientifiques (IHES); 35 route de
Chartres; F--91440 {\bf Bures-sur-Yvette}, FRANCE.}\\

   {Physics Department, McGill University\\
   3600 University Street, Montr\'eal, Qu\'ebec, CANADA, H3A  2T8}\\
   {and}\\
   {Laboratoire de physique nucl\'eaire, Universit\'e
   de Montr\'eal\\
   CP 6128, Succ centre-ville; Montr\'eal, Qu\'ebec, CANADA,
   H3C  3J7}\\
\vskip 2truecm

\end{center}
\vfill\eject   
\begin{abstract}
{We consider an embedding of the extremal four-dimensional 
Reissner-Nordstr\"om black hole into type $IIB$ string theory. 
The equivalent type $IIB$ configuration, in the D-brane weak-coupling
picture, is a bound state of D1- and D5-branes threaded by fundamental 
type $IIB$ strings. The bound state involves also a NSNS solitonic 
5-brane, mimicking the role of the Kaluza-Klein magnetic monopole. 
The statistical entropy derived by counting the degeneracy of the 
BPS-saturated excitations of this bound state agrees perfectly with 
the (semiclassical) Bekenstein-Hawking formula.\\}  
\end{abstract}
 }
\end{titlepage}

\setcounter{footnote}{0}
\def\beq{\begin{equation}}
\def\eeq{\end{equation}}
\def\beqn{\begin{eqnarray}}
\def\barrl{\begin{array}{ll}}
\def\eeqn{\end{eqnarray}}
\def\earr{\end{array}}


\setcounter{footnote}{0}
\section{Introduction}
In the early seventies Hawking showed~[1] 
using classical general relativity and quantum 
field theory in curved spacetime that black holes 
radiate thermally at a temperature 
$T_H = \kappa / 2 \pi$, where $\kappa$ is the surface 
gravity. Since the presence of temperature always 
calls for an entropy the back hole acquires one 
which is equal to one quarter of the horizon 
area~[1,2]. With this discovery, the earlier beautiful
(but mere) analogy between the black hole classical mechanics 
and the laws of thermodynamics were put onto a much 
firmer foundation. Since thermodynamics is only a 
low-energy approximation to a more fundamental 
(microscopic) description, there ought to be a 
statistical interpretation of the black hole 
entropy in terms of its fundamental underlying 
quantum degrees of freedom. Over the past 
years, there have been different attempts to solve
this problem and they all agree that a satisfactory 
and complete resolution of the problem requires a 
full quantum theory of gravity not just a semiclassical 
approximation~[3--13].
\smallskip

Recently, there has been a considerable progress 
in this direction. This came through our improved
understanding of the Ramond-Ramond (RR) charged 
string solitons through the Dirichlet brane (D-brane) 
description~[14--20].
This rapid progress has led to a microscopic state 
counting of the black hole entropy in an explicit 
string computation using the D-brane technology~[21].
\smallskip

The results of the D-brane counting of the black 
hole entropy were curiously all limited to five
dimensions at the beginning~[21--24]. 
The reason for this was that suitable 
four-dimensional black holes with non-zero horizon 
which are bound states of D-branes alone were not 
known\footnote{The situation has changed since then. 
Recently, extremal four-dimensional black holes 
with finite area were constructed solely from 
intersecting D-branes of type II on $T^6$~[25]. 
These systems of D-branes
can also be interpreted in terms of intersecting brane
configurations of $\M$-theory~[26]. In the latter case,
one may use the 5-brane low-energy effective theory from 
quantization of the boundary states of the $\M$-theory 
two branes ending on the 5-branes to give the BPS
state counting of the entropy~[27].}.
One needs to add to the weak 
coupling bound state equivalent (of the semiclassical 
four-dimensional black holes) a Kaluza-Klein monopole 
like in the work of~[28] or a symmetric 5-brane 
that carries an axion charge as performed in the 
work of~[29]. The presence of the Kaluza-Klein monopole's 
magnetic charge (or the symmetric 5-brane's axion charge) 
will ensure that the four-dimensional string coupling 
is constant everywhere. This is a necessary condition 
for any D-brane counting of the Bekenstein-Hawking 
entropy to work~[25,30]. 
\smallskip

It is the purpose of our work here to study another
example of these four-dimensional extremal black holes
with non-vanishing horizon area and show, using the newly
discovered D-brane technology, that its Bekenstein-Hawking 
entropy precisely matches the statistical entropy arising from 
the degeneracy of the BPS excitations (at weak coupling) of 
the corresponding D-brane bound state. Our four-dimensional 
extremal black hole is given by its supersymmetric embedding 
into a six-dimensional type $IIB$  string theory compactified 
on $K3$. We find that the section $\left(t , r , \theta , 
\varphi\right)$ of the resulting black hole looks like a 
four-dimensional extremal Reissner-Nordstr\"om solution 
where the extra fifth $(x^4)$ and sixth $(x^5)$ coordinates
are wrapping ${\widehat S}{}^1$ and $S{}^1$, respectively. 
In~[28] two six-dimensional black hole examples 
were presented as uplifted four-dimensional extremal 
Reissner-Nordstr\"om charged black hole solutions of the 
Einstein-Maxwell gravity. One of the solutions solves
the background field equations of the type $IIA$ 
six-dimensional string theory on $K3$. The other one 
is a solution of the type $IIB$ six-dimensional string theory 
on $K3$. From the discussion of~[28] one
can use the $T-$duality transformation along the $x^5$ 
direction to map the type $IIA$ and type $IIB$ solutions 
problems to one another. The six-dimensional embedding 
that we deal with in the present calculation is a 
$T-$dualised version along the $x^4$ direction of the
type $IIA$ solution of~[28]. We find that the equivalent
D-brane bound-state problem involves a solitonic NSNS 
5-brane mimicking the role of the Kaluza-Klein monopole 
in stabilizing the horizon and making the four-dimensional
dilaton (and hence the string coupling) constant.
\smallskip

In Section~(2), we shall describe our six-dimensional
solution as an uplifted four-dimensional extremal 
Reissner-Nordstr\"om black hole, and argue for its 
being supersymmetric. In Section~(3), we compute 
the ADM mass, the horizon area and the Hawking 
temperature (which is zero at extremality). We also 
display the different RR and NSNS charges that are 
involved and express the entropy in terms of them. 
Section~(4) deals with the statistical
entropy from counting the BPS-saturated massless
excitations of the of the weak coupling bound state 
which corresponds to the strong coupling extremal black hole 
configuration found in Section~(2). The agreement with the
Bekenstein-Hawking entropy is again remarkable.
Our presentation ends in Section~(5) by some concluding 
remarks and directions for future investigations.    
\smallskip

\section{The making of the extremal four-dimensional 
black hole}
The six-dimensional type $IIB$ superstring action 
including compactification on $K3$ 
space is given in the string-frame metric
by\footnote{We could also consider compactification
on $T^4$. We shall use the same notation as in~[28] 
and units where $\alpha' = 1$.}
\begin{equation}
S_{IIB} = {8\, V_4}\, \int\, d^6\!x\, \sqrt{-g}\,
\left[ e^{-2\phi}\left(R + 4\, (\partial\phi)^2
- (\partial\sigma)^2 - {1\over{12}}\, H^2\right)
- {e^{2\sigma}\over{12}}\, F_{(3)}^2\right].
\label{aa}
\end{equation}
Apart from $F_{(3)}$, the field content of this type 
$IIB$ action is common to the other 6-dimensional superstring 
theories, $IIA$ on $K3$ and heterotic on the four-Torus $T^4$. 
The field $\sigma$ is the modulus scalar for $K3$ and $\phi$ 
is the six-dimensional dilaton. 
$H$ is the 3-form field strength for the NSNS 
2-form $B$, and $F_{(3)}$ is the 3-form field 
strength for the RR 2-form $A_{(2)}$. The six-dimensional 
solution that solves the background field equations of 
action $S_{IIB}$ is given by:
\begin{eqnarray}
ds^2 & = & -\, f^{-2}\, dt^2 + f^2\, 
\left(dr^2 + r^2 d\theta^2 + 
r^2\sin^2\theta d\varphi^2\right)
\nonumber\\
& & +\; {1\over{4\, k^2}}\, \left(dx^4\right)^2
+ {1\over{L^2}}\, \left(dx^5 \right)^2 \;;
\nonumber\\
A_{(2)} & = & {L\over{\la\lt}}\, f^{-1}\, dt \wedge dx^4 +
{{2\, k^2}\over{\la\lt}}\, \cos\theta\, d\varphi \wedge dx^5\; ;
\nonumber\\
B & = & {1\over L}\, \left(f^{-1} - 1\right)\, dt \wedge dx^5
+ {1\over 2}\; \cos\theta\, d\varphi \wedge dx^4 \;;
\nonumber\\
e^\phi  & = & {{\la}\over{2\, L\, k}}\;;\qquad 
e^\sigma=\lt\;;\qquad f = 1 + {k \over r}\;.
\label{ab}
\end{eqnarray}
The $\left(t, r, \theta, \varphi\right)$ subspace 
of solution (\ref{ab}) yields the four-dimensional 
extremal Reissner-Nordstr\"om black hole.
The extra dimensions $x^4$ and $x^5$ are
compactified on ${\widehat S}{}^1$ and $S{}^1$,
respectively.
It is also worth noting that the subspace
$\left(\theta,\varphi, x^4\;\, \hbox{or}\;\, x^5\right)$
has now the topology of $S{}^2\times 
\left({\widehat S}{}^1\;\, \hbox{or}\;\, S{}^1\right)$.
This observation will in fact be useful later 
(in Section~(3)) when we come to calculate the 
NSNS and RR charges generated by $B$ and $A_{(2)}$ 
2-forms as well as the Bekenstein-Hawking entropy. 
We also notice that the magnetic charge replacing 
the Kaluza-Klein monopole of~[28,29] 
is now given by  the component $B_{\varphi 4}$ of 
the NSNS 2-form $B_{\mu\nu}$. 
\smallskip

Following the discussion of~[28],
the quick way to see that the solution in
(\ref{ab}) is supersymmetric and preserves
$1/4$ of the spacetime supersymmetries of
the $N=2$ ten-dimensional theory
is to find its ten-dimensional up-lift.
However, one of the conclusions drawn from the investigation
of~[31] (and which are essential for our
supersymmetry discussion) is the realization
that the truncation of the type $IIB$
string theory to a ten-dimensional $N=1$ theory
containing only its NSNS fields\footnote{These NSNS 
fields are the usual NSNS 2-form $B$ generating the 
3-form field strength $H$ and a ten-dimensional 
NSNS 2-form $A'_{(2)}$ generating a 3-form field 
strength $F'_{(3)}$ not to be confused with RR 2-form 
$A_{(2)}$ of (\ref{aa}).} admits the extremal 
Reissner-Nordstr\"om black hole as solution. Most notably, 
this solution is supersymmetric and preserves 
$1/4$ of the $N=1$, $D=10$ theory, and is related to
the ten-dimensional up-lift of (\ref{ab}) under the
interchange of the RR and NSNS 2-forms. Furthermore, 
since in (\ref{aa}) the action $S_{IIB}$ has it 4-form 
potential set to zero the RR and NSNS 3-form field strengths
should enter symmetrically in the supersymmetry transformations 
of the $N=1$, $D=10$ type $IIB$ theory~[32].
Consequently, the ten-dimensional up-lift of the 
RR solution in (\ref{ab}) should mimic the NSNS
Reissner-Nordstr\"om solution of~[31],
and preserves $1/4$ of the $N=1$, $D=10$ spacetime
supersymmetries. The compactification to six-dimensions
yields back the solution (\ref{ab}) which is then seen 
to preserve $1/4$ of the spacetime supersymmetries of the
$N=2$ theory.  
\smallskip

Next, we move to computing the different charges 
and thermodynamic quantities of the solution (\ref{ab}).
\smallskip

\section{Charges and thermodynamic quantities}
The extremal black hole solution in (\ref{ab})
can carry both electric and magnetic charges with 
respect to the RR 3-form field strength $F_{(3)}$: 
\begin{eqnarray}
Q_1 & = & {{V_4}\over{2\, \pi^{5/2}}}\;
\int_{S{}^2\times S{}^1}\;\; e^{2\sigma}\; {}^* F_{(3)} =
{{2^3}\over{\sqrt{\pi}}}\; V_4\; {\lt\over\la}\; k^2\; ;
\nonumber\\
Q_5 & = & 2^3\, \pi^{3/2}\;
\int_{S{}^2\times S{}^1}\;\; F_{(3)} =
2^7\, \pi^{7/2}\; {{k^2}\over{\lt\la}}\; .
\label{af}
\end{eqnarray}
It carries also the following fundamental `electric'
and solitonic `magnetic' charges with respect to the 
NSNS 3-form field strength $H$:
\begin{eqnarray}
W_F & = & 2^4\,\pi\, V_4\;
\int_{S{}^2\times {\widehat S}{}^1}\;\;
e^{-2\phi}\; {}^* H =
2^8\, \pi^3\, V_4\, {{k^2\, L^2}\over{{\la^2}}}\;;
\nonumber\\
W_S & = & {1\over{{2^2}\, \pi^2}}\; 
\int_{S{}^2\times {\widehat S}{}^1}\;\; H =  
 1 \;.
\label{ag}
\end{eqnarray}    
The ADM mass of solution (\ref{ab}) 
is\footnote{For the calculation of the ADM mass 
we have used the definition given in.~[33].}
\begin{equation}
M_{ADM} = 2^{10}\, \pi^3\, V_4\, 
{{k^2 L}\over{\la^2}}\; .
\label{ah}
\end{equation}
The vanishing of $g_{t\mu} = 0$ indicates that 
the solution carries no ADM momentum and the 
Hawking temperature is given by
\begin{equation}
T_H = {1\over{2\, \pi}}\; \left|{1\over f}\; \partial_r\, 
f^{-1}\right|_{\hbox{at the horizon}}\; ,
\label{ai}
\end{equation}
which typically vanishes for the extremal black hole 
configuration of (\ref{ab}) at the horizon $r=0$.
Using the six-dimensional metric and the knowledge that
the topology of the horizon is  
$S{}^2\times {\widehat S{}^1}\times S{}^1$,
the horizon area of the six-dimensional extremal 
black hole is then given by
\begin{equation}
\A = 2^3\, \pi^3\, {k\over L}\; .
\label{aj}
\end{equation}
The Bekenstein-Hawking entropy is then simply
\begin{equation}
S = {\A\over{\left(4 G_N\right)}} = 2^{10}\, \pi^4\, V_4\, 
{{k^3 L}\over{\la^2}}\; ,
\label{ak}
\end{equation}
where we have used that the Newton's constant
$1/{4\, G_N} = 2^7\, \pi\, V_4\, k^2\, L^2/\la^2$.
In terms of the charges carried by the black hole
the entropy $S$ is easily seen to take the much
simpler form
\begin{equation}
S = 2\pi\; \sqrt{Q_1\, Q_5\, W_F\, W_S} = 
2\pi\; \sqrt{Q_1\, Q_5\, W_F} \; .
\label{al}
\end{equation}
This alternative way of expressing the 
Bekenstein-Hawking entropy will (in the next 
section) turn out to be useful when we come to 
compare it with the statistical entropy derived 
by using the D-brane counting method. A noteworthy 
feature also of equation (\ref{al}) is that the solitonic 
charge enters trivially in the entropy since $W_S =1$. 
We shall appreciate more this point when
we come to present the D-brane picture in Section~(4).
\smallskip

\section{The statistical entropy from the D-brane picture}
In this section we describe the weak coupling bound 
state problem to which we map the semiclassical
black hole field configuration (\ref{ab}). In the weak 
coupling problem, we compute the statistical 
entropy of a collection of D-branes and strings and
reproduce the Bekenstein-Hawking formula (\ref{al}).
\smallskip

\subsection{The constituents of the weak coupling bound state}
Before dealing with the weak coupling problem, let us
review the elements of the strong coupling problem 
at hand. From the discussion of Section~(3), the
strong coupling problem is 
an extremal black hole solution carrying $Q_1$
electric charge and $Q_5$ magnetic charge with 
respect to the RR $3-$form field strength.
This black hole exists also in the background field
of macroscopic fundamental strings and a solitonic 
5-brane carrying respectively the $W_F$ electric 
charge and unit $W_S$ of magnetic charge with respect to 
the NSNS 2-form $B$.
\smallskip

In turning to the description of the weak coupling
equivalent configuration, we should recall 
that in the usual formulation of type $IIB$ string theory 
we find the RR $p-$form field strengths, for $p = 3, 5, 7$. 
Due now to the celebrated work of Polchinski~[16], we 
understand these different RR $p-$forms as having as sources
the D1-, D3- and D5-branes, respectively.
In a $\sigma-$model description, we say that in
ten dimensions the world volume of the Dp-branes
couple to the type $IIB$ RR $(p+1)-$form potential
$A_{(p+1)}$, and hence they carry electric charge 
with respect to the RR $(p+2)-$form field strength
$F_{(p+2)}$. Therefore, to identify the weak coupling 
D-brane bound state (to which we map the configuration 
(\ref{ab})) one has simply to count the total number
of charges from the RR fields in the solution (\ref{ab}).
Since solution (\ref{ab}) involves also electric and
magnetic $B$ charges, we are sure that the bound 
state is in the presence of fundamental type $IIB$
strings wrapping $x^5$ and a NSNS solitonic 
5-brane wrapping $K3\times x^5$.
\smallskip

So based on what precedes, our weak coupling bound 
state is made of $Q_1$ D1-branes wrapping themselves 
around ${\widehat S}{}^1$ (or $x^4$) and carrying 
each a unit electric RR 3-form charge. Their partner in the D-brane 
bound state is $Q_5$ D5-branes which carry each a unit magnetic RR 
3-form charge and wrap the entire 
$K3\times {\widehat S}{}^1$ space. 
The latter will in fact appear as strings in six dimensions,
therefore forming a D-string composite with the
$Q_1$ D1-branes\footnote{The normalizations in (\ref{af})
and (\ref{ag}) are defined in such a way that the charges 
$Q_1$, $Q_5$, $W_F$ and $W_S$ are naturally integer 
quantized. Thus the amount of each of the RR charges involved 
in (\ref{ab}) count the actual number of the D-branes 
entering in the bound state structure.}. Owing to the 
presence of the NSNS 2-form $B$, the complete
picture of the bound state is a D-string composite
threaded by fundamental type $IIB$ strings
winding around $S{}^1$ and carrying the
NSNS 3-form electric charge $W_F$ 
(which plays the role of a winding number). 
There is also the NSNS solitonic 5-brane 
carrying the unit $W_S$ magnetic charge with respect 
$H$ and winding around the whole of the internal 
space $K3\times S{}^1$, and hence overlap
the $Q_5$ D5-branes along $K3$.
The BPS excitations of this weak bound state 
will preserve the supersymmetries of the $N=2$ theory 
as does its semiclassical strong coupling equivalent.
\smallskip

\subsection{The microscopic counting of entropy}
For the purpose of computing the entropy of the
semiclassical extremal black hole, we shall now
simply count the BPS-saturated excitations of the
underlying bound state configuration just 
described~[21]. Our counting method
is actually similar to the description in~[29,34,35].
The D-string composite at hand consists of $Q_5$ parallel 
D5-branes wrapping $K3\times {\widehat S}{}^1$ , 
$Q_1$ parallel D1-branes wrapping ${\widehat S}{}^1$ ,
and total $W_F$ winding number along 
$x^5\,\left(S{}^1\right)$, which is carried 
by the fundamental type $IIB$ strings. Let us consider
the limit of 
large\footnote{The D-brane counting in the 
case where all the charges $Q_1$, $Q_5$ and $W_F$ 
are of the same order is discussed in~[36].}  
$W_F$ and $Q_1$. Since duality
implies that entropy can only depend on the
product\footnote{The work of~[18--20] 
explicitly checks this in some special 
cases.} $Q_1\, Q_5$ it is sufficient to do the counting
for the case\footnote{For other values of $Q_5$
the counting problem of the BPS-saturated string states
may be different but duality relates the resulting entropies.
For instance if $Q_5 > 1$, see the discussion below which 
uses a counting method in which
the bound state of a D1- and D5-branes is pictured as an
instanton of the $U({Q_5})$ gauge theory on the D5-brane.} 
$Q_5 = 1$. The $Q_1$ D1-branes are then marginally bound to the
D5-brane but are free to wander within the transverse
$K3$ space. This motion is generated by $\left(1,1\right)$
Dirichlet open strings both of whose ends are stuck to the
D1-branes. (The $\left(1,5\right)$
Dirichlet open strings do not contribute 
in our (particular $Q_5 = 1$) counting problem
due to charge confinement~[29,34,35].)  
This yields $4 Q_1 Q_5$ underlying winding energy 
microstates with minimum mass in boson-fermion pairs 
available within the D-string composite which generate 
the transverse motion.
The BPS-saturated excitations of the bound state 
are actually these $4 Q_1 Q_5$ boson-fermion
pairs counted as right-moving $\left(1,1\right)$ 
Dirichlet open strings 
(with no left-moving Dirichlet open strings) 
that end on the D1-branes 
and sharing the total winding number $W_F$ going 
around $x^5\,\left(S{}^1\right)$.
\smallskip

An alternative way to count these BPS-saturated string 
states is to picture the D1-branes as instantons
of the $U\left( Q_5 \right)$ gauge theory on
the worldvolume of the D5-brane~[37,17]. More precisely,
the D1-branes are the zero size limit of these instantons.
In the work of~[20], it was shown that the moduli space
of one of these instantons (one D1-brane) corresponds 
to an $N=4$ superconformal field theory with $4\, Q_5$ 
bosonic and an equal number of fermionic degrees of 
freedom. For a number of $Q_1$ D1-branes, 
the degrees of freedom of the resulting 
``instanton strings''  are parametrized by 
$4\, Q_1\, Q_5$ physical oscillators in boson-fermion
pairs\footnote{A counting method along these lines
was recently presented in a different but related
context~[38]. We are thankful to
J. Maldacena for pointing this to us.}.
The logarithm of the maximum number of ways we distribute 
$W_F$ amongst these string states gives the 
statistical entropy that we are after and which from 
the standard $(1+1)-$effective field theory on the string 
is given by the formula:
\begin{equation}
S = \sqrt{{\pi\, \left(2 {N_B} + N_F\right)\, 
E_R\, \L}\over{6}}\; ,
\label{am}
\end{equation}
where $N_B\, \left(N_F\right)$ is the number of species
of right-moving bosons (fermions) with $E_R$ total 
winding energy\footnote{The winding energy is a `stringy'
property that does not exist in field theory. In order
to wrap around the circle the string must be stretched,
in so doing its energy increases with the winding number. This
contribution is of course to be added to the usual one
coming from the center-of-mass momentum.} 
in a box of length $\L$. Using
$N_B = N_F = 4 Q_1 Q_5$ and $E_R = 2\pi W_F /\L$,
we find (for the large $W_F$ thermodynamic limit)
\begin{equation}
S_{stat} = 2\pi\, \sqrt{Q_1\, Q_5\, W_F}\; ,
\label{an}
\end{equation}     
this is in prefect agreement with the 
Bekenstein-Hawking entropy of the semiclassical 
black hole (\ref{al}).
\smallskip

Since in our problem (and from (\ref{ag})) we have $W_S = 1$, 
the D-brane counting presented in the preceding discussion
is not affected by the presence of the single
NSNS solitonic 5-brane. We should recall
(in similarity with the $Q_1$ and $Q_5$ RR charges)
that the charge normalization in (\ref{ag}) are chosen in 
such a way that the charge $W_S$ count the actual number
of the NSNS solitonic 5-branes.  Had we had $W_S > 1$,
we would have had these $W_S$ 5-branes (which can
be located anywhere on ${\widehat S}{}^1$ ) 
intersecting all the $Q_1$ 
D1-branes\footnote{When $Q_5 = 1$ we have 
seen that the only relevant Dirichlet open 
strings for the counting of the degeneracy of 
the BPS-saturated excitations of the 
bound state are of $\left(1,1\right)$ type, whose 
ends are stuck on the $Q_1$ D1-branes.}. 
As a result, the $Q_1$ closed D1-branes break up 
into $\left(W_S\, Q_1\right)$ open 1-branes, each 
of which is bounded by a pair of the NSNS solitonic 
5-branes. Hence, the winding number carried by 
the Dirichlet $\left(1,1\right)$ open strings 
will carry an additional label referring to 
the pair of the NSNS 5-branes they lie in 
between~[29]. The number of the underlying
microstates in boson-fermion pairs becomes
$N_B = N_F = 4\, Q_1\, Q_5\, W_S$. Inserting
this into (\ref{am}) with $E = 2\pi W_F /\L$ we
obtain
\begin{equation}
S_{stat} = 2\pi\, \sqrt{Q_1\, Q_5\, W_F\, W_S}\; ,
\label{ao}
\end{equation}     
in agreement with (\ref{an}) if $W_S =1$. 
\smallskip

Finally, we recall that in our problem 
the NSNS solitonic 5-brane plays
the role of the Kaluza-Klein monopole~[28] 
in maintaining a constant modulus 
for the $x^4$ direction, and hence keeping 
the four-dimensional string coupling constant. 
A quantitative way to check this is to write down 
the equations of motion for the modulus field 
$\sigma$ and the dilaton $\phi$ from the the type 
$IIB$ action (\ref{aa}) after reformulating 
it in the Einstein-frame metric, 
$g_{E\mu\nu} = e^{-\,\phi}\, g_{S\mu\nu}$. 
A glance at their equations of motion reveals 
that in order for $\sigma$ and $\phi$ to be constant 
one has to have the $B_{\varphi 4}$ component of 
the NSNS $B-$field, which yields the magnetic charge 
$W_S$ carried by the NSNS solitonic 5-branes.
\smallskip

\section{Concluding remarks}
We have displayed a new example of embedding the four-dimensional
extremal Reissner-Nordstr\"om black hole into type $IIB$
string theory. The solution carries both electric and
magnetic charges with respect to the RR 3-form field
strength $F_{(3)}$ and the NSNS 3-form field strength
$H$. The corresponding weak coupling D-brane bound state
involves D1-branes winding around $x^4$ along with
D5-branes wrapping $K3\times {\widehat S{}^1}$. 
The resulting D-string composite is threaded by 
fundamental type $IIB$ strings winding around $x^5$. 
There is also a NSNS solitonic 5-brane wrapping 
$K^3\times S{}^1$ whose role is to 
replace the Kaluza-Klein monopole in
maintaining the $\phi$ and $\sigma$ fields constant.
\smallskip

We have also given an explicit derivation of 
the statistical entropy of the underlying weak 
coupling bound state configuration by counting 
the number of its BPS-saturated excitations
(arising as Dirichlet low-energy open string states 
stuck on the D-branes). By exploiting the virtues of 
an $N=4$ or $N=8$ supersymmetric theory, one is able 
to extrapolate the result of the weak coupling limit 
using the D-brane picture back to the strong coupling 
regime. It is satisfying to find that the D-brane 
calculation gives an entropy which continues to match 
perfectly well the semiclassical Bekenstein-Hawking formula.
\smallskip

It would be interesting to study more four-dimensional
black holes, as they would potentially shed more light
on the counting problem of the statistical entropy
from the D-brane bound states in string theory. 
A new class of background solutions can be found, for example,
by using a combined $S\,T$ duality in six dimensions~[32]
to map type $IIB$ on $K3$ to heterotic string theory on
$T^4$, then performing a standard $O(22,6)$ rotation
and using the $S\,T$ duality afterwards to go back to
type $IIB$ on $K3$ string theory. An interesting calculation
would be then to find the weak coupling D-brane bound state
equivalents to the new backgrounds thus obtained, and
provide for each example the counting of the 
microscopic states responsible for the statistical 
entropy. This type of investigation would 
hopefully help in providing a more general approach to
the entropy problem, which covers any four-dimensional
black hole background solution.
\smallskip

At the end, may be, we should point out again that
we have been restricted so far only to extreme and 
slightly near-extreme black hole configurations. 
This arises since in the present stage of the D-brane 
technology one can only count states at weak coupling, 
whereas black holes only exist at strong coupling. 
There appears to be no justification why should a 
weakly coupled description continue to hold far 
from extremality. However, most recently it was shown 
that there is a sense in which even black holes far 
away from extremality can be mapped into a composite
of weakly interacting fundamental strings and 
D-branes\footnote{There have been earlier indications 
in~[22] that the counting of string states at weak 
coupling agrees with the black hole entropy even
in situations where one could not justify the 
extrapolation to strong coupling.}.
This question was addressed in~[34] for the 
five-dimensional black holes and generalized to
the four-dimensional case in~[35]. It would
certainly be interesting to extend their work 
to other four-dimensional examples and also see 
whether the symmetry of the resulting statistical 
entropy continues to be consistent with $U-$duality
symmetry.
\bigskip
\bigskip
\bigskip

\centerline{\bf Acknowledgments}
First I would like to thank the Laboratoire 
de Physique Nucl\'eaire (Universit\'e
de Montr\'eal) for their hospitality 
while this research was carried out.
It is a pleasure to express all my gratitude to 
R.~T.~Sharp for his continuous support, help and 
encouragement. Finally, I would like also to thank
R.~C.~Myers, J.~Maldacena, R.~R.~Khuri, M.~Paranjape 
and R.~T.~Sharp for useful discussions. This research 
was supported by NSERC of Canada and Fonds FCAR du 
Qu\'ebec.
\bigskip

\section{References}

\noindent
[1]. 
S. Hawking, Nature 248 (1974) 30;

\noindent
\phantom{[1].}
S. Hawking Comm. Math. Phys. {\bf 43} (1975) 199.
\medskip

\noindent 
[2]. 
J. Bekenstein, Lett. Nuov. Cimento {\bf 4} (1972) 737;

\noindent 
\phantom{[2].} 
S. Hawking Phys. Rev. {\bf D7} (1973) 2333;

\noindent 
\phantom{[2].} 
S. Hawking Phys. Rev. {\bf D9} (1974) 3292.
\medskip

\noindent
[3]. 
J. Bekenstein, Phys. Rev. {\bf D12} (1975) 3077.
\medskip

\noindent
[4].
S. Hawking, Phys. Rev {\bf D13} (1976) 191.
\medskip

\noindent
[5].
W. Zurek and K. Thorne, Phys. Rev. Lett. {\bf 54} 
(1985) 2171.
\medskip

\noindent
[6].
G. 't Hooft, Nucl. Phys. {\bf B335} (1990) 138;

\noindent 
\phantom{[6].} 
G. 't Hooft Phys. Scr. {\bf T36} (1991) 247.
\medskip

\noindent
[7].
L. Susskind, {\bf hep-th/9309145}.
\medskip

\noindent
[8].
L. Susskind and J. Uglum, Phys. Rev. {\bf D50} (1994) 2700 
{\bf [hep-th/9401070]}.
\medskip

\noindent
[9]. 
C. Teitelboim, {\bf hep-th/9510180}.
\medskip

\noindent
[10]. 
A. Sen, Mod. Phys. Lett. {\bf A10} (1995) 2081 
{\bf [hep-th/9504147]}.
\medskip

\noindent
[11]. 
S. Carlip, {\bf gr-qc/9509024}.
\medskip

\noindent
[12]. 
F. Larsen and F. Wilkczek, {\bf hep-th/9511064}.
\medskip

\noindent
[13]. 
M. Cvetic and A. Tseytlin, {\bf hep-th/9512031}.
\medskip

\noindent
[14]. 
J. Dai, R. G. Leigh and J. Polchinski, 
Mod. Phys. Lett. {\bf A4} (1989) 2073.
\medskip

\noindent
[15]. 
P. Horava, Phys. Lett. {\bf B231} (1989) 251. 
\medskip

\noindent
[16]. 
J. Polchinski, Phys. Rev. Lett. {\bf 75} (1995) 4724 
{\bf [hepth/9510017]}.

\noindent
\phantom{[16].}
J.~Polchinski, S.~Chaudhuri and C.~V.~Johnson,
{\bf hep-th/9602052}.
\medskip

\noindent
[17]. 
E. Witten, {\bf hep-th/9510135}.
\medskip

\noindent
[18]. 
A. Sen, {\bf hep-th/9510229} and 
{\bf hep-th/9511026}.
\medskip

\noindent
[19]. 
M. Bershadsky, V. Sadov and C. Vafa, 
{\bf hep-th/9511222}.
\medskip

\noindent
[20]. 
C. Vafa, {\bf hep-th/9511088} and 
{\bf hep-th/9512078}.
\medskip 

\noindent
[21]. 
A. Strominger and C. Vafa, {\bf hep-th/9601029}.
\medskip

\noindent
[22].
 C. Callan and J. Maldacena, {\bf hep-th/9602043}.
\medskip

\noindent
[23]. 
G. Horwitz and A. Strominger, {\bf hep-th/9602051}.
\medskip

\noindent
[24]. 
J. C. Breckenridge, R. C. Myers, A. W. Peet
and C. Vafa, {\bf hep-th/9602065};

\noindent 
\phantom{[24].} 
J.~C.~Breckenridge, D.~A.~Lowe, R.~C.~Myers, A.~W.~Peet,
A.~Strominger and C.~Vafa, {\bf hep-th/9603078}.
\medskip 

\noindent
[25]. 
V. Balasubramanian and F. Larsen, {\bf hep-th/9604189}.
\medskip

\noindent 
[26]. 
I. R. Klebanov and A. A. Tseytlin, {\bf hep-th/9604166}.
\medskip 

\noindent 
[27]. 
R.~Dijkgraaf, E.~Verlinde and H.~Verlinde,
{\bf hep-th/9603126}.

\noindent
\phantom{[27].}
R.~Dijkgraaf, E.~Verlinde and H.~Verlinde,
{\bf hep-th/9604055}.
\medskip 

\noindent
[28]. 
C. V. Johnson, R. R. Khuri and R. C. Myers,
{\bf hep-th/9603061}.
\medskip

\noindent
[29].
J. M. Maldacena and A. Strominger,
{\bf hep-th/9603060}.
\medskip

\noindent
[30].
F. Larsen and F. Wilczek, {\bf hep-th/9511064}.

\noindent
\phantom{[30]}
S. Ferrara and R. Kallosh, {\bf hep-th/9603090}.
\medskip

\noindent
[31].
R.~R.~Khuri and T.~Ort\'{\i}n, {\bf hep-th/9512177};
{\bf hep-th/9512178}.
\medskip

\noindent
[32]. 
E. Bergshoeff, C. M. Hull and T. Ort\'{\i}n,
Nucl. Phys. {\bf B451} (1995) 547 {\bf [hep-th/9504081]};

\noindent 
\phantom{[32].} 
K.~Behrndt, E.~Bergshoeff and B.~Janssen,
{\bf hep-th/9512152}.
\medskip

\noindent
[33].
J. X. Lu, Phys. Lett. {\bf B313} (1993) 29.

\noindent 
\phantom{[33].} 
R. R. Khuri and R. C. Myers, {\bf hep-th/9512061}.
\medskip

\noindent
[34].
G. T. Horwitz, J. M. Maldacena and A. Strominger,
{\bf hep-th/9603109}.
\medskip

\noindent
[35].
G. T. Horwitz, D. A. Lowe and J. M. Maldacena,
{\bf hep-th 9603195}.
\medskip

\noindent
[36].
J. Maldacena and L. Susskind, {\bf hep-th/9604042}.
\medskip

\noindent
[37].
M. Douglas, {\bf hep-th/9512077}.
\medskip

\noindent
[38].
J. Maldacena, {\bf hep-th/9605016}.
\medskip

\end{document}